\documentclass[aps,prb,twocolumn,groupedaddress,floats,showpacs]{revtex4}
\usepackage{latexsym}
\usepackage{dcolumn}
\usepackage[dvips]{graphicx}
\usepackage{amssymb}
\usepackage{graphics}
\usepackage{amsmath}
\usepackage{epsf}

\begin{document}
\author{S. Adam and S. Das Sarma}
\title{Transport in suspended graphene}
\affiliation{Condensed Matter Theory Center, Department of Physics, 
University of Maryland, College Park, MD 20742-4111, USA}
\date{\today}
\begin{abstract}
Motivated by recent experiments on suspended graphene showing carrier 
mobilities as high as $200,000~{\rm cm}^{2}/Vs$, we theoretically
calculate transport properties assuming Coulomb impurities 
as the dominant scattering mechanism.  We argue that
the substrate-free experiments done in the diffusive regime
are consistent with our theory and verify 
many of
our earlier predictions including (i) removal of the substrate will increase
mobility since most of the charged impurities are in the substrate, 
(ii) the minimum conductivity is not universal, but depends on 
impurity concentration with cleaner samples having a higher minimum 
conductivity.  We further argue that experiments on suspended graphene
put strong constraints on the two parameters involved in our theory,
namely, the charged impurity concentration $n_{\rm imp}$ and $d$, the typical 
distance of a charged impurity from the graphene sheet.  The recent
experiments on suspended graphene indicate a residual impurity
density of $1-2 \times 10^{10}~{\rm cm}^{-2}$ which are presumably 
stuck to the graphene interface, compared to impurity densities
of $\sim 10^{12}~{\rm cm}^{-2}$ for graphene on SiO$_2$ substrate.  
Transport experiments can therefore be used as a spectroscopic tool 
to identify the properties of the remaining impurities in suspended 
graphene.       
\end{abstract}
\pacs{81.05.Uw; 72.10.-d, 73.40.-c}
\maketitle

\section{Introduction}
\label{Sect:Intro}

The recent experimental realization of a single layer of carbon atoms
arranged in a honeycomb lattice has prompted much excitement in both
theoretical and experimental physics communities (For a recent
review, see Ref.~\onlinecite{kn:dassarma2007a} and references
therein).  Most experiments so far have been performed on a SiO$_2$
substrate using the mechanical exfoliation technique developed in
Ref.~\onlinecite{kn:novoselov2004} followed by optical detection that
is sensitive to the thickness of the substrate.  In these experiments
one expects there to be charged impurities located in the substrate
with a typical density of $n_{\rm imp} \sim 10^{12}~{\rm cm}^{-2}$
with the impurities sitting near the substrate-graphene interface with
a typical distance $d\lesssim 1~{\rm nm}$ from the graphene sheet.  In
previous work~\cite{kn:hwang2006c,kn:adam2007a} we argued that
removing the substrate would remove the charged impurities and enhance
mobility.  Other models of scattering such as ripples, on the other
hand, would lead to enhanced scattering in suspended graphene thereby
reducing the mobility, since a freely suspended graphene sheet is
expected to be more rippled than one glued to a substrate.

Very recently experiments have been performed on suspended
graphene~\cite{kn:bolotin2008,kn:du2008} that find remarkably high
mobilities $\mu = 200,000~{\rm cm}^{2}/Vs$ for carrier transport
confirming that charged impurities in the substrate were indeed the
dominant scattering mechanism limiting the mobility.  Signatures from 
these new experiments (such as the conductivity still being linear in carrier
density) imply that the remaining impurities in suspended graphene are
long-range and are probably charged dopants lying (i.e. stuck) on the graphene
surface, or remnants of chemicals used during the preparation.  The
minimal model we propose here contains only two parameters: the charged
impurity density and the typical distance of the impurity from the
graphene sheet.  Theoretical estimates for these numbers in suspended
graphene are hard to come by, for example, a recent DFT
study~\cite{kn:rytkonen2007} argues that potassium and sodium dopants
are typically located at $d\sim 0.3~{\rm nm}$ from a graphite surface.
In what follows we treat $n_{\rm imp}$ and $d$ as parameters, and argue
that the experimental signatures both at high density and low density
effectively constrain these numbers.  Turning the problem around,
the recent transport experiments on suspended graphene could be used
as a spectroscopic tool to identify the properties of the remaining
impurities in suspended graphene.  For example, a mobility
$\mu = 200,000~{\rm cm}^{2}/Vs$, implies that $n_{\rm imp} = 1.73 \times 10^{10}
~{\rm cm}^{-2}$, or equivalently, that the distance between the remaining 
impurities is around $76~{\rm nm}$ with the charged impurities
sitting typically at an effective  distance of $0.1 - 1~{\rm nm}$ from the graphene
sheet.

\section{Theoretical Model}
\label{Sect:Theory}
To calculate the transport properties of graphene, we consider
impurities with an effective $2D$ density $n_{\rm imp}$ located 
in a plane at a distance $d$, either above or below the graphene sheet.  
While in reality, different
impurities may be located at different distances, one should therefore
think of $d$ as an effective or typical impurity distance 
and $n_{\rm imp}$ as an average 2D impurity density. Alternatively, one 
could imagine several impurity planes at different
distances $d_i$, and then adding up their respective contributions to 
conductivity using Matthiessen's rule.  As we show below, impurities located 
farther away are exponentially suppressed, so in this respect, one could 
then think of $d$ then as distance of the closest charged impurities and 
thus
dominating the transport properties.  In any case, our minimal model
for charged impurities contains only two parameters $n_{\rm imp}$ and $d$.  

For an impurity located at the origin, the bare Coulomb potential at the point
${\bf r}$ in the graphene sheet is
\begin{equation}
\phi_0({\bf r}) = \frac{Z e^2}{\kappa} \frac{1}{\sqrt{r^2 + d^2}},
\label{Eq:Coulomb}
\end{equation}
where $Ze$ is the impurity charge and $\kappa$ is an effective background
lattice dielectric constant.  All the properties of the substrate 
(or vacuum in the case of suspended graphene) can be absorbed into a third
parameter $r_s = Ze^2/\kappa \gamma$, where $\gamma/\hbar = v_{\rm F}
\approx c/300$.  Here $c$ is the speed of light, and $v_{\rm F}$ is 
the constant graphene Fermi velocity that relates energy to momentum in the
linear Dirac spectrum $ \varepsilon_{\rm F} = \gamma k_{\rm F}$.
The carrier density $n = k_{\rm F}^2/\pi$.  Although we 
keep $r_s$ as an arbitrary number in the theory, we note for
concreteness that $r_s = 0.8$ for graphene on a Si0$_2$ substrate,
and $r_s = 2$ for suspended graphene (see 
Refs.~\onlinecite{kn:adam2007a,kn:hwang2006c} for further details on the 
model).  The Fourier transform of Eq.~\ref{Eq:Coulomb} gives
\begin{eqnarray}
{\tilde \phi}_0(q) = 2 \pi \gamma r_s \frac{ e^{-q d}}{q}, \nonumber \\
{\tilde \phi}(q) = 2 \pi \gamma r_s \frac{ e^{-q d}}{\epsilon(q) q},
\end{eqnarray}
where in the second line, we assume linear screening with
\begin{equation}
\epsilon(q) = \left\{ \begin{array}{ll} 
                        1 + 4 k_{\rm F} r_s/q & \ \ q \leq 2 k_{\rm F} \\
                        1 + \pi r_s/2        & \ \ q > 2 k_{\rm F}. 
\end{array} \right.
\end{equation} 
Although our choice of dielectric function has an artificial
discontinuity at $q = 2 k_{\rm F}$, its advantage is that it allows
for analytic results, and we have checked by numerical integration
that this choice of dielectric function and the Random Phase
Approximation~\cite{kn:ando2006,kn:hwang2006b} give identical results for all 
graphene transport
properties (see Ref.~\onlinecite{kn:adam2007a} and references therein
for a discussion of this point).  Further, we observe that the chiral
properties induced by the sub-lattice structure of the honeycomb
lattice imply that in contrast to most other electronic systems,
$2k_{\rm F}$ scattering is absent in graphene.

Graphene conductivity can then be calculated in Boltzmann transport
theory where $\sigma = 2 (e^2/h) \varepsilon_{\rm F}
\tau_{\varepsilon_{\rm F}}$ and
\begin{equation}
\frac{\hbar}{\tau(\varepsilon_k)} = 2 \pi  \sum_{k'} n_{\rm imp}
         |{\tilde \phi}(|{\bf k} - {\bf k}'|)|^2 
         (1 - \hat{{\bf k}}\cdot \hat{{\bf k}}')
         |\zeta_{\bf k'}^\dagger \zeta_{\bf k}|^2  
\delta(\varepsilon_k - \varepsilon_{k'}), \nonumber
\end{equation}
where $\zeta_{\bf k}$ is the spinor component of the electron (or hole) 
eigenvector of the $2\times2$ Dirac Hamiltonian.  These expressions
can be simplified by introducing two dimensionless variables
$\eta = q/2 k_{\rm F}$ and $a = 4 k_{\rm F} d = 4 \pi^{1/2} d \sqrt{n}$ 
to find
\begin{eqnarray}
\sigma &=& \frac{e^2}{h} \left( \frac{n}{n_{\rm imp}} \right) 
\frac{1}{I(a, r_s)}, \nonumber \\
I(a, r_s) &=& 2 r_s^2 \int_0^1 d\eta \eta^2 \sqrt{1-\eta^2} 
\frac{e^{-a \eta}}{(\eta + 2 r_s)^2}
\label{Eq:Integral}
\end{eqnarray}

To calculate the transport properties of graphene from Coulomb
impurities the integral $I(a, r_s)$ was solved numerically by
Ando~\cite{kn:ando2006}, Cheianov and Falko~\cite{kn:cheianov2006} and
Hwang et al~\cite{kn:hwang2006c}.  We explore several limits in which
$I(a, r_s)$ can be solved analytically.  In the ``Complete Screening''
limit, $d\rightarrow 0, r_s \rightarrow \infty$, Nomura and
MacDonald~\cite{kn:nomura2006a} showed that $I(a, r_s \rightarrow
\infty)= \pi/32$. This result does not capture the effect of changing
the substrate which in turn changes the parameter $r_s$.  This
complete screening approximation is therefore inadequate to study
transport in suspended graphene even at a qualitative level.  For
example, as we show below, keeping $n_{\rm imp}$ the same, but just
changing the dielectric constant from SiO$_2$ to vacuum, would
actually decrease mobility by almost a factor of 2.  Therefore, in
suspended graphene the two effects of reducing both dielectric
constant and $n_{\rm imp}$ actually compete to determine graphene
mobility (see also Ref.~\onlinecite{kn:sabio2007}).  Taking
$d\rightarrow 0$, but keeping the $r_s$ dependence, we
showed~\cite{kn:adam2007a} that $I(0, r_s) = 2/G(x=2r_s)$
\begin{equation}
\frac{G[x]}{x^2} = \frac{\pi}{4} + 3x - \frac{3 \pi x^2}{2} 
+ \frac{x(3x^2 - 2)\arccos[1/x]}{\sqrt{x^2 -1}}.
\end{equation}
We note here that the following additional limits can be 
solved: $I(a, r_s \rightarrow 0) = (r_s^2 \pi/a) (I_1(a) + L_1(a))$,
where $I_1$ is a modified Bessel function and $L_1$ is a modified
Struve function.  We also have $I(a \rightarrow \infty, r_s) \sim a^{-3}$, 
where this last result says that for very large density, Coulomb impurities
in graphene give $\sigma(n \gg 1) \sim n^{2.5}$.  However, this is not 
the limit relevant to experiments, which are all in the low density regime.
The following expansion captures both the substrate (i.e. $r_s$ dependence) 
and the $d$ dependence, and is relevant for current experiments both on 
suspended graphene and for graphene on a SiO$_2$ substrate
\begin{equation}
\sigma = 2 \frac{e^2}{h} \frac{1}{G[x] - d \sqrt{n} F[x]},
\label{Eq:Main}
\end{equation}     
where
\begin{equation}
\frac{F[x]}{2 \sqrt{\pi} x^2} = \frac{1}{3} - \frac{\pi x}{2}
 - 4 x^2 + 2 \pi x^3 + \frac{x^2(3-4x^2)\arccos[1/x]}{\sqrt{x^2 -1}}.
\end{equation} 
We note that for most relevant substrates $x>1$, (recall that 
for a SiO$_2$ substrate $x = 2 r_s \approx 1.6$ and for
suspended graphene $x \approx 4$), and that the 
ratio $A[x] = F[x]/G[x]$ does not change significantly from
its large $r_s$ asymptote of $A \rightarrow 128/(15 \sqrt{\pi})$.  Therefore 
for practical comparisons with experiments, one can 
approximate graphene conductivity using
\begin{equation}
\sigma(n) \approx \frac{2 e^2}{h} \left(\frac{n}{G[2 r_s] n_{\rm imp}}
\right) \frac{1}{1 - d \sqrt{n} \frac{128}{15 \sqrt{\pi}}}.
\label{Eq:Useful}
\end{equation} 
For SiO$_2$ substrate $G[2 \times 0.8] \approx 0.1$, giving the $d=0$ result
of $\sigma \approx 20 (e^2/h)(n/n_{\rm imp})$, and for suspended
graphene, $G[2\times 2] \approx 0.144$ giving $\sigma \approx 13.86
(e^2/h)(n/n_{\rm imp})$.  We find Eq.~\ref{Eq:Useful} to be a
very useful formula for all existing bulk graphene transport
data in the literature.  

We now turn to the discussion of mobility $\mu$.  Defining mobility
in graphene is both important and problematic.  The importance stems
mostly from it being the figure of merit for graphene's application
as a practical transistor device.  However, close to the Dirac point
we have $V_g \rightarrow 0$ but finite resistivity.  If one 
naively assumes that $n \sim V_g$ in this regime {\it as is often
done}, then one can get arbitrarily high values of mobility which is
completely meaningless (a detailed discussion of this point can be
found in Ref.~\onlinecite{kn:hwang2006e}).  Calculating mobility at
very high density is complicated by the fact that
both  super-linear contributions (see Eq.~\ref{Eq:Main} for finite $d$) and 
sub-linear contributions from short-range scatterers 
(see Ref.~\onlinecite{kn:adam2007a,kn:hwang2006c} and discussion in 
Sect.~\ref{Sect:Discuss} below) come into play, so that
mobility is not constant for a given sample.  In this case,
mobility would be a non-monotonic function of carrier density whose
maximum value is given by a competition of different scattering 
mechanisms present in graphene.  We have consistently
argued that the best definition of graphene mobility is the slope 
$d\sigma/dn$ just beyond the Dirac point conductivity plateau, 
i.e. only within the
window of density where the conductivity is {\it strictly linear} in $V_g$.  We
advocate this choice because at low density, one can show
theoretically that both the sub-linear and super-linear contributions
to the conductivity are small.  For this definition of mobility, one
can immediately see that
\begin{equation}
\mu [m^2/Vs] = \frac{5}{G[x]}\frac{n_0}{n_{\rm imp}},
\label{Eq:mob}
\end{equation}
where $n_0 = 10^{10} {\rm cm}^{-2}$.  Although our theory has two free
parameters, i.e. $n_{\rm imp}$ and $d$, knowing the mobility immediately
determines $n_{\rm imp}$ through Eq.~\ref{Eq:mob}.  A careful
characterization of the super-linear contribution using
Eq.~\ref{Eq:Main} could then be used to determine $d$.  In addition,
as we describe below, the low density transport properties i.e. the
value of the minimum conductivity, the width of the minimum
conductivity plateau and the offset of the Dirac point depend only on
these two parameters that in principle could already be determined
from the high-density transport data.         
      
So far we have formulated everything in terms of carrier 
density $n = k_{\rm F}^2/\pi$.  For large external 
gate voltage, we can map our results to applied gate voltage 
using $n = \alpha V_g$, where $\alpha$ is experimentally 
determined to be $7.2 \times 10^{10} {\rm cm}^{-2}{\rm V}^{-1}$ for 
graphene on SiO$_2$ and $3.76 \times 10^{10} {\rm cm}^{-2}{\rm V}^{-1}$ 
for suspended graphene~\cite{kn:tan2007,kn:bolotin2008}.  The last step is 
to determine the carrier density $n$ close to the Dirac point.  The low density
conductivity arises from the carriers in the charge-impurity induced 
puddles of electrons and holes~\cite{kn:hwang2006c}.  We argued
in Ref.~\onlinecite{kn:adam2007a} that at low density it was appropriate to 
use the {\it rms} density caused by the inhomogeneous potential 
induced by the same charged impurities that are responsible for
the high-density transport properties.  A self-consistent calculation
gives~\cite{kn:adam2007a}
\begin{eqnarray}
\frac{n_{\rm rms}}{n_{\rm imp}} &=& \frac{x^2}{2} 
\left( -1 + \frac{16 E_1[a]}{(4 + \pi x)^2} \right.  \\
&& \mbox{} \left. +\frac{e^{-a}x}{1+x} 
  + (1+x)e^{ax}(E_1[ax] - E_1[a(1+x)]) \right), \nonumber 
\end{eqnarray}     
where $E_1(z) = \int_{z}^{\infty} t^{-1} \exp^{-t} dt$ is the
exponential integral function and the self-consistency is obtained by
setting $a = 4 d \sqrt{\pi n_{\rm rms}}$.  The offset in the Dirac point
is given by ${\bar n} = n_{\rm imp}^2/4 n_{\rm rms}$ (a derivation of
this result can be found in Ref.~\onlinecite{kn:adam2007a}). The value
of $n_{\rm rms}$ calculated from this self-consistent calculation
agrees with a recent calculation that minimizes the graphene energy
functional in the presence of disorder~\cite{kn:rossi2008} and for $d
\sim 1~{\rm nm}$ is consistent with the experimental observations of
density inhomogeneities on a SiO$_2$ substrate~\cite{kn:martin2007}.
Given the choice of substrate (which determines $r_s$) and picking
$n_{\rm imp}$ and $d$ completely determines the conductivity at high
density and the non-universal disorder dependent minimum conductivity
plateau at low density.

\section{Results}
\label{Sect:Results}
\begin{figure}
\bigskip
\epsfxsize=0.9\hsize
\epsffile{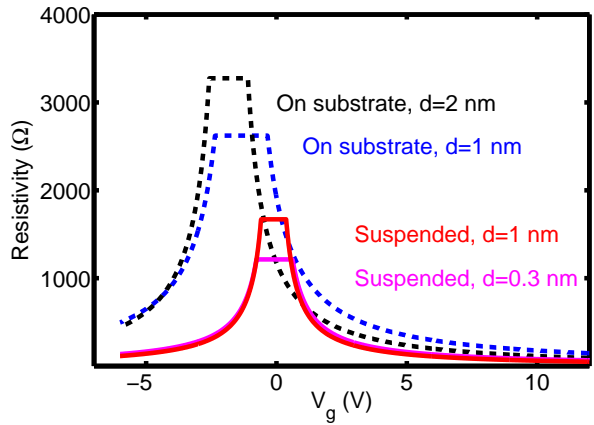}
\caption{\label{Fig:rho} (Color online) Transport properties
of suspended graphene (solid lines) compared to graphene on 
a SiO$_2$ substrate (dashed lines).  The self-consistent carrier
density theory predicts that the cleaner suspended graphene samples
would have a higher minimum conductivity (i.e. lower maximum 
resistivity), smaller off-sets and narrower
plateaus.  As described in the text, we use 
$n_{\rm imp}(SiO_2) = 1.7 \times 10^{11}~{\rm cm}^{-2}$ (i.e. 
mobility $\mu = 30,000~{\rm cm}^{2}/Vs$) for graphene on a substrate
and $n_{\rm imp} = 1.7 \times 10^{10}~{\rm cm}^{-2}$ (
$\mu = 200,000~{\rm cm}^{2}/Vs$) for substrate-free graphene.}
\end{figure}

To illustrate our results, in what follows, we choose the experimental
parameters reported in Ref.~\onlinecite{kn:bolotin2008}.  We use
mobility $\mu = 30,000~{\rm cm}^{2}/Vs$ for graphene on a substrate
(corresponding to $n_{\rm imp} \approx 1.7\times 10^{11}~{\rm
cm}^{-2}$) and $\mu = 200,000~{\rm cm}^{2}/Vs$ for suspended graphene
($n_{\rm imp} \approx 1.7\times 10^{10}~{\rm cm}^{-2}$).  We also
employ a hard cut-off between the low and high density regime such
that $n = n_{\rm rms}$ if $\alpha V_g < n_{\rm rms}$ and $n= \alpha
V_g$ for $\alpha V_g > n_{\rm rms}$.  Shown in Fig.~\ref{Fig:rho} is
the resistivity as a function of gate voltage for both suspended
graphene and graphene on a SiO$_2$ substrate.  An important
observation is that the higher mobility samples have a lower
resistivity which is an important prediction of our self-consistent
theory~\cite{kn:adam2007a}.  This has been verified both in transport
measurements on SiO$_2$ substrate~\cite{kn:tan2007} and by
intentionally adding charged impurities in ultra-high
vacuum~\cite{kn:chen2007b}.  The recent data~\cite{kn:bolotin2008} on 
ultra-high mobility suspended graphene samples further confirm that 
charged impurities provide the dominant scattering mechanism in
graphene and are responsible for the {\it non-universal} minimum
conductivity.  We note that in plotting resistivity, all the
super-linear behavior is obscured and the high-density curves of the
same mobility but different $d$ look quite similar.  By contrast, from
the properties of the low density ``Dirac plateau'' one can
extract the magnitude of $d$, but we caution that we only expect our
self-consistent carrier density theory to agree within a factor of two
with the experimentally observed plateau width, Dirac point offset and 
the maximum resistivity.  A careful analysis of the experimental data 
should be able to extract the best values for $n_{\rm imp}$ and $d$ as has 
been done previously for graphene on a 
substrate~\cite{kn:tan2007,kn:chen2007b}.    
         
\begin{figure}
\bigskip
\epsfxsize=0.9\hsize
\epsffile{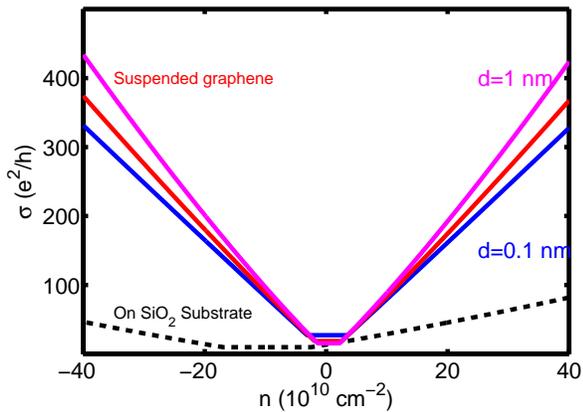}
\caption{\label{Fig:Cond} (Color online) Suspended graphene conductivity
using the same values of $n_{\rm imp}$ as in Fig.~\protect{\ref{Fig:rho}}
and $d = 0.1, 0.5$, and $1~{\rm nm}$.  Plotting the conductivity instead of 
resistivity reveals some additional properties including the predicted 
super-linear behavior shown in the figure that can be used to determine 
the distance of the charged impurities from the graphene sheet.}
\end{figure}

\begin{figure}
\bigskip
\epsfxsize=0.9\hsize
\epsffile{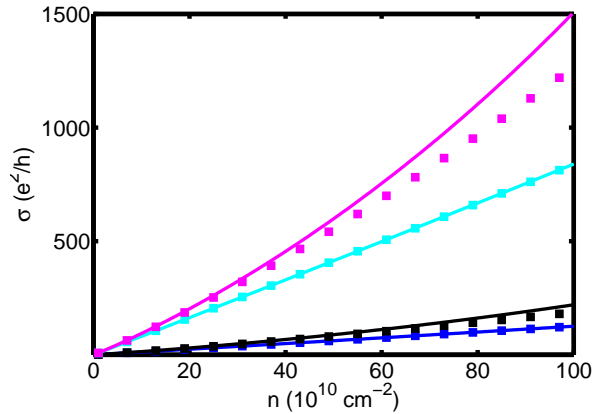}
\caption{\label{Fig:Verify} (Color online) Comparison 
of analytical result Eq.~\protect{\ref{Eq:Main}} (solid line)
and a numerical evaluation of Eq.~\protect{\ref{Eq:Integral}} for
realistic parameters in suspended graphene and on a substrate.  The
curves from top to bottom are suspended graphene with $d=1~{\rm nm}$
and $d=0.1~{\rm nm}$, and graphene on a SiO$_2$ substrate with 
 $d=1~{\rm nm}$ and $d=0.1~{\rm nm}$ respectively.  The values
of $n_{\rm imp}$ are the same as in Fig.~\protect{\ref{Fig:rho}}.}
\end{figure}

In Fig.~\ref{Fig:Cond} we show conductivity as a function of ``carrier
density'' defined through $n=\alpha Vg$.  For experimentally 
reasonable carrier densities, one can notice that $d=1~{\rm nm}$
and $d=0.1~{\rm nm}$ should be distinguishable by the super-linear
contributions of the former.  In many ways plotting the
data as conductivity instead of resistivity reveals much more 
information.  From these results we argue that transport data not only
provide information about $n_{\rm imp}$ but also $d$, thereby being a tool
to not only identify different types of scattering mechanisms (e.g. 
short-range, long-range, phonons) but for Coulomb impurities, it 
provides information about the location of the impurities. In 
Fig.~\ref{Fig:Verify} we show our analytical 
expression Eq.~\ref{Eq:Main} (solid lines) compared to 
a numerical evaluation of Eq.~\ref{Eq:Integral} (squares) for
different values of $d$ and for both suspended graphene and on 
a substrate.  One finds that for most of the experimental regime,
the analytical expression is an excellent approximation for 
the Boltzmann conductivity, but that at very high-density and large $d$, 
it slightly over-estimates the super-linear behavior.  However, in 
this high-density regime, one also expects the contribution of 
non-Coulombic
short-range scatterers which reduce the conductivity.  For large $d$ and 
a particular ratio of short-range and Coulomb scatterers, one may even have 
non-monotonic mobility that first increases because of our predicted
super-linear contribution and then decreases as short-range scattering 
becomes more dominant at higher densities.

\begin{figure}
\bigskip
\epsfxsize=0.9\hsize
\epsffile{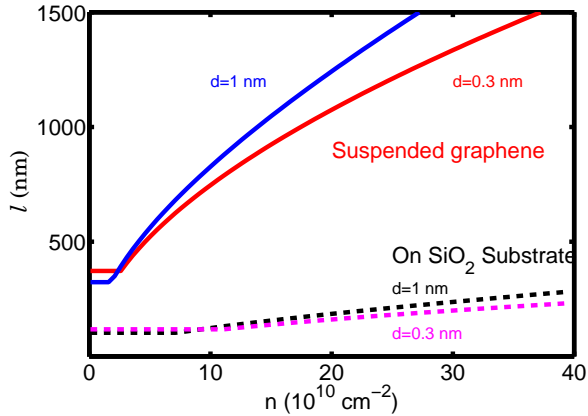}
\caption{\label{Fig:mfp} (Color online) Mean free
path $\ell$ for suspended graphene (solid lines) and on a substrate
(dashed lines) using the same parameters as in
Fig.~\protect{\ref{Fig:rho}}.  For sample sizes $L \sim 1 \mu m$, we
observe that close to the Dirac point, transport for both suspended
graphene and on a substrate is in the diffusive regime, where our
theory applies.  For larger carrier densities, suspended graphene
crosses over to a ballistic transport regime, whereas for graphene on
a substrate the transport is always diffusive.}
\end{figure}

Finally, in Fig.~\ref{Fig:mfp}, we show our calculated
carrier mean free path ($\ell$) as a function of carrier 
density, comparing graphene on SiO$_2$ substrate with 
suspended graphene.  It is remarkable that, for 
reasonable parameters (same as Fig.~\ref{Fig:rho}) used
in these results, the suspended graphene $\ell > 1.5~\mu{\rm m}$ 
for $n\sim 3 \times 10^{11}~{\rm cm}^{-2}$
whereas for the same density range in graphene on 
SiO$_2$ substrate, the charged impurities
give rise to an almost constant $\ell \sim 0.1-0.2~\mu{\rm m}$.
As noted by Bolotin et al.~\cite{kn:bolotin2008} such 
long ($>1 \mu{\rm m}$) mean free paths are comparable to the 2D
system size of the samples, implying ballistic transport
and a breakdown of our diffusive theory.  It is important
to realize that for mean free paths comparable to the sample 
size, strong short-range scattering will be imposed by the sample
edges leading to a saturation of the 
conductivity and the mean-free-paths around this density.  Note,
however, that the current graphene samples on the substrate
are {\it always} in the diffusive limit, and graphene
transport around the Dirac point (both with and without substrate)
is also {\it always} in the diffusive limit since the mean free path 
is much shorter than the sample size.  Of course, by making
nanoribbons~\cite{kn:miao2007,kn:han2007} with sample sizes
comparable to $\ell$, one can artificially induce ballistic
transport.

\section{Discussion}
\label{Sect:Discuss}
In this work, we have generalized the transport theory of 
Ref.~\onlinecite{kn:adam2007a} to the case of suspended
graphene~\cite{kn:bolotin2008} and derived analytical results for 
the super-linear 
contribution to conductivity expected for Coulomb scatterers at
a finite distance $d$ from the graphene sheet.  However, this 
effect becomes pronounced only at high density where short-range
scattering begins to dominate over Coulomb scattering.  We note
that in the experiments of Bolotin et. al~\cite{kn:bolotin2008},
they were only able to reach rather low carrier densities, and 
therefore we believe that Coulomb scattering dominates
over all other scattering mechanisms in their results.
However, as was pointed
out in Ref.~\onlinecite{kn:bolotin2008}, there may be strong
short-range scattering by sample edges when the mean free path becomes
comparable to the system size.  However one can anticipate 
for the Dirac plateau in cleaner samples (such as future suspended
graphene experiments), charged impurities will induce a smaller
carrier density $n_{\rm rms}$, {\it even as the ratio} $n_{\rm
rms}/n_{\rm imp} \sim \sigma$ is increasing.  For example, a sample
with mobility of $\mu = 500,000~{\rm cm}^2/Vs$, would have $n_{\rm
imp} = 0.7 \times 10^{10}~{\rm cm}^{-2}$ which for $d=0.3~{\rm nm}$
gives $n_{\rm rms} \approx 1.2 \times 10^{10}~{\rm cm}^{-2}$ and
$\sigma_{min} \approx 20 e^2/h$.  This gives $\ell \approx 600~{\rm
nm}$.  So despite the high-mobility, the peculiar scaling $\ell \sim
k_{\rm F}$ for Coulomb scatterers in graphene implies that close to
the charge neutral point, we are in the diffusive Boltzmann transport
regime.  However, at higher carrier density we have $\ell \approx
22~k_{\rm F}/n_{\rm imp}$, suggesting that the ballistic regime could
be explored away from the Dirac point.  For mesoscopic samples
constructed in the geometry of $\ell \gtrsim L$, where $L$ is the
sample size, our theory does not apply.  

Smaller sample sizes would also increase the effect of edge
scattering, which act as short-range scattering centers.  As was
argued in Hwang et al.~\cite{kn:hwang2006c} at high density these
begin to dominate showing a non-universal cross-over from the
low-density regime where $\sigma$ is linear in $n$ to the high-density
limit where $\sigma$ is constant.  So while in this work we have focused
on Coulomb scattering that dominates at low density, probing a wider
density range could also distinguish between different scattering 
mechanisms.  We emphasize here that detailed measurements
and analysis of $\sigma(n)$ could thus provide in-depth 
spectroscopic information about the impurity distribution in graphene,
far surpassing the resolution of currently available scanning
microscopy measurements~\cite{kn:martin2007} which directly probe
the potential landscape.

Finally we note that alternative theories for the dominant scattering
mechanism in graphene make predictions that are inconsistent with the
results of Bolotin et al.~\cite{kn:bolotin2008}.  For example, if
ripples were the dominant scattering mechanism, then suspended
graphene would have stronger ripples and hence lower mobility.  It was
one of our early predictions that removing charged impurities in the
substrate would increase mobility and increase the minimum
conductivity.  The initial conventional wisdom in the theoretical
community that graphene experiments were observing the universal Dirac
minimum conductivity should now be discarded after recent experimental
developments including intentionally doping graphene with potassium in
Ref.~\onlinecite{kn:chen2007b} and measurements done on graphene with
mobility varying close to three orders of magnitude in
Refs.~\onlinecite{kn:tan2007,kn:bolotin2008} and most particularly the
suspended graphene measurements, all of which clearly demonstrate the
importance of charged impurity disorder controlling the minimum
conductivity plateau.  It was in
Refs.~\onlinecite{kn:hwang2006c,kn:adam2007a} that we predicted that
charged impurities in graphene would give rise to a non-universal
minimum conductivity whose value would increase with cleaner samples.
These predictions have now been vindicated in light of recent
experiments.  It, therefore, appears that the existing graphene
transport experiments are completely dominated by charged impurity
scattering, and no universal ``Dirac point physics'' is in play near
the charge neutrality point.  The physics of the charge neutrality point
is dominated by diffusive transport through electron and hole puddles
induced by the charged impurities.  We believe that the close agreement 
between our theory and the transport data in 
suspended graphene experiments establishes the dominance of charged
impurity scattering in graphene beyond any reasonable doubt. 

This work is supported by U.S. ONR and NSF-NRI.  

\vspace{0.2in}

\begin{thebibliography}{19}
\expandafter\ifx\csname natexlab\endcsname\relax\def\natexlab#1{#1}\fi
\expandafter\ifx\csname bibnamefont\endcsname\relax
  \def\bibnamefont#1{#1}\fi
\expandafter\ifx\csname bibfnamefont\endcsname\relax
  \def\bibfnamefont#1{#1}\fi
\expandafter\ifx\csname citenamefont\endcsname\relax
  \def\citenamefont#1{#1}\fi
\expandafter\ifx\csname url\endcsname\relax
  \def\url#1{\texttt{#1}}\fi
\expandafter\ifx\csname urlprefix\endcsname\relax\def\urlprefix{URL }\fi
\providecommand{\bibinfo}[2]{#2}
\providecommand{\eprint}[2][]{\url{#2}}

\bibitem[{\citenamefont{\mbox{Das Sarma} et~al.}(2007)\citenamefont{\mbox{Das
  Sarma}, Geim, Kim, and MacDonald}}]{kn:dassarma2007a}
\bibinfo{editor}{\bibfnamefont{S.}~\bibnamefont{\mbox{Das Sarma}}},
  \bibinfo{editor}{\bibfnamefont{A.~K.} \bibnamefont{Geim}},
  \bibinfo{editor}{\bibfnamefont{P.}~\bibnamefont{Kim}}, \bibnamefont{and}
  \bibinfo{editor}{\bibfnamefont{A.~H.} \bibnamefont{MacDonald}}, eds.,
  \emph{\bibinfo{title}{Exploring Graphene: Recent Research Advances, A Special
  Issue of Solid State Communications}}, vol. \bibinfo{volume}{143}
  (\bibinfo{publisher}{Elsevier}, \bibinfo{year}{2007}).

\bibitem[{\citenamefont{Novoselov et~al.}(2004)\citenamefont{Novoselov, Geim,
  Morozov, Jiang, Zhang, Dubonos, Grigorieva, and Firsov}}]{kn:novoselov2004}
\bibinfo{author}{\bibfnamefont{K.~S.} \bibnamefont{Novoselov}},
  \bibinfo{author}{\bibfnamefont{A.~K.} \bibnamefont{Geim}},
  \bibinfo{author}{\bibfnamefont{S.~V.} \bibnamefont{Morozov}},
  \bibinfo{author}{\bibfnamefont{D.}~\bibnamefont{Jiang}},
  \bibinfo{author}{\bibfnamefont{Y.}~\bibnamefont{Zhang}},
  \bibinfo{author}{\bibfnamefont{S.~V.} \bibnamefont{Dubonos}},
  \bibinfo{author}{\bibfnamefont{I.~V.} \bibnamefont{Grigorieva}},
  \bibnamefont{and} \bibinfo{author}{\bibfnamefont{A.~A.}
  \bibnamefont{Firsov}}, \bibinfo{journal}{Science}
  \textbf{\bibinfo{volume}{306}}, \bibinfo{pages}{666} (\bibinfo{year}{2004}).

\bibitem[{\citenamefont{Adam et~al.}(2007)\citenamefont{Adam, Hwang, Galitski,
  and {\mbox Das Sarma}}}]{kn:adam2007a}
\bibinfo{author}{\bibfnamefont{S.}~\bibnamefont{Adam}},
  \bibinfo{author}{\bibfnamefont{E.~H.} \bibnamefont{Hwang}},
  \bibinfo{author}{\bibfnamefont{V.~M.} \bibnamefont{Galitski}},
  \bibnamefont{and} \bibinfo{author}{\bibfnamefont{S.}~\bibnamefont{{\mbox Das
  Sarma}}}, \bibinfo{journal}{Proc. Natl. Acad. Sci. USA 104, 18392}
  (\bibinfo{year}{2007}).

\bibitem[{\citenamefont{Hwang et~al.}(2007{\natexlab{a}})\citenamefont{Hwang,
  Adam, and \mbox{Das Sarma}}}]{kn:hwang2006c}
\bibinfo{author}{\bibfnamefont{E.~H.} \bibnamefont{Hwang}},
  \bibinfo{author}{\bibfnamefont{S.}~\bibnamefont{Adam}}, \bibnamefont{and}
  \bibinfo{author}{\bibfnamefont{S.}~\bibnamefont{\mbox{Das Sarma}}},
  \bibinfo{journal}{Phys. Rev. Lett.} \textbf{\bibinfo{volume}{98}},
  \bibinfo{eid}{186806} (\bibinfo{year}{2007}{\natexlab{a}}).

\bibitem[{\citenamefont{Bolotin et~al.}(2008)\citenamefont{Bolotin, Sikes,
  Jiang, Fudenberg, Hone, Kim, and Stormer}}]{kn:bolotin2008}
\bibinfo{author}{\bibfnamefont{K.}~\bibnamefont{Bolotin}},
  \bibinfo{author}{\bibfnamefont{K.}~\bibnamefont{Sikes}},
  \bibinfo{author}{\bibfnamefont{Z.}~\bibnamefont{Jiang}},
  \bibinfo{author}{\bibfnamefont{G.}~\bibnamefont{Fudenberg}},
  \bibinfo{author}{\bibfnamefont{J.}~\bibnamefont{Hone}},
  \bibinfo{author}{\bibfnamefont{P.}~\bibnamefont{Kim}}, \bibnamefont{and}
  \bibinfo{author}{\bibfnamefont{H.}~\bibnamefont{Stormer}},
  \bibinfo{journal}{arXiv:0802.2389v1}  (\bibinfo{year}{2008}).

\bibitem[{\citenamefont{Du et~al.}(2008)\citenamefont{Du, Skachko, Barker, and
  Andrei}}]{kn:du2008}
\bibinfo{author}{\bibfnamefont{X.}~\bibnamefont{Du}},
  \bibinfo{author}{\bibfnamefont{I.}~\bibnamefont{Skachko}},
  \bibinfo{author}{\bibfnamefont{A.}~\bibnamefont{Barker}}, \bibnamefont{and}
  \bibinfo{author}{\bibfnamefont{E.}~\bibnamefont{Andrei}},
  \bibinfo{journal}{arXiv:0802.2933v1}  (\bibinfo{year}{2008}).

\bibitem[{\citenamefont{Rytk\"{o}nen et~al.}(2007)\citenamefont{Rytk\"{o}nen,
  Akola, and Manninen}}]{kn:rytkonen2007}
\bibinfo{author}{\bibfnamefont{K.}~\bibnamefont{Rytk\"{o}nen}},
  \bibinfo{author}{\bibfnamefont{J.}~\bibnamefont{Akola}}, \bibnamefont{and}
  \bibinfo{author}{\bibfnamefont{M.}~\bibnamefont{Manninen}},
  \bibinfo{journal}{Phys. Rev. B} \textbf{\bibinfo{volume}{75}},
  \bibinfo{pages}{075401} (\bibinfo{year}{2007}).

\bibitem[{\citenamefont{Hwang and \mbox{Das Sarma}}(2007)}]{kn:hwang2006b}
\bibinfo{author}{\bibfnamefont{E.~H.} \bibnamefont{Hwang}} \bibnamefont{and}
  \bibinfo{author}{\bibfnamefont{S.}~\bibnamefont{\mbox{Das Sarma}}},
  \bibinfo{journal}{Phys. Rev. B} \textbf{\bibinfo{volume}{75}},
  \bibinfo{pages}{205418} (\bibinfo{year}{2007}).

\bibitem[{\citenamefont{Ando}(2006)}]{kn:ando2006}
\bibinfo{author}{\bibfnamefont{T.}~\bibnamefont{Ando}}, \bibinfo{journal}{J.
  Phys. Soc. Jpn.} \textbf{\bibinfo{volume}{75}}, \bibinfo{pages}{074716}
  (\bibinfo{year}{2006}).

\bibitem[{\citenamefont{Cheianov and Fal'ko}(2006)}]{kn:cheianov2006}
\bibinfo{author}{\bibfnamefont{V.}~\bibnamefont{Cheianov}} \bibnamefont{and}
  \bibinfo{author}{\bibfnamefont{V.}~\bibnamefont{Fal'ko}},
  \bibinfo{journal}{Phys. Rev. Lett.} \textbf{\bibinfo{volume}{97}},
  \bibinfo{pages}{226801} (\bibinfo{year}{2006}).

\bibitem[{\citenamefont{Nomura and MacDonald}(2006)}]{kn:nomura2006a}
\bibinfo{author}{\bibfnamefont{K.}~\bibnamefont{Nomura}} \bibnamefont{and}
  \bibinfo{author}{\bibfnamefont{A.~H.} \bibnamefont{MacDonald}},
  \bibinfo{journal}{Phys. Rev. Lett.} \textbf{\bibinfo{volume}{96}},
  \bibinfo{pages}{256602} (\bibinfo{year}{2006}).

\bibitem[{\citenamefont{Sabio et~al.}(2007)\citenamefont{Sabio, Seoanez,
  Fratini, Guinea, \mbox{Castro Neto}, and Sols}}]{kn:sabio2007}
\bibinfo{author}{\bibfnamefont{J.}~\bibnamefont{Sabio}},
  \bibinfo{author}{\bibfnamefont{C.}~\bibnamefont{Seoanez}},
  \bibinfo{author}{\bibfnamefont{S.}~\bibnamefont{Fratini}},
  \bibinfo{author}{\bibfnamefont{F.}~\bibnamefont{Guinea}},
  \bibinfo{author}{\bibfnamefont{A.~H.} \bibnamefont{\mbox{Castro Neto}}},
  \bibnamefont{and} \bibinfo{author}{\bibfnamefont{F.}~\bibnamefont{Sols}},
  \bibinfo{journal}{arXiv:0712.2232v3}  (\bibinfo{year}{2007}).

\bibitem[{\citenamefont{Hwang et~al.}(2007{\natexlab{b}})\citenamefont{Hwang,
  Adam, and \mbox{Das Sarma}}}]{kn:hwang2006e}
\bibinfo{author}{\bibfnamefont{E.~H.} \bibnamefont{Hwang}},
  \bibinfo{author}{\bibfnamefont{S.}~\bibnamefont{Adam}}, \bibnamefont{and}
  \bibinfo{author}{\bibfnamefont{S.}~\bibnamefont{\mbox{Das Sarma}}},
  \bibinfo{journal}{Phys. Rev. B} \textbf{\bibinfo{volume}{76}},
  \bibinfo{pages}{195421} (\bibinfo{year}{2007}{\natexlab{b}}).

\bibitem[{\citenamefont{Tan et~al.}(2007)\citenamefont{Tan, Zhang, Bolotin,
  Zhao, Adam, Hwang, \mbox{Das Sarma}, Stormer, and Kim}}]{kn:tan2007}
\bibinfo{author}{\bibfnamefont{Y.-W.} \bibnamefont{Tan}},
  \bibinfo{author}{\bibfnamefont{Y.}~\bibnamefont{Zhang}},
  \bibinfo{author}{\bibfnamefont{K.}~\bibnamefont{Bolotin}},
  \bibinfo{author}{\bibfnamefont{Y.}~\bibnamefont{Zhao}},
  \bibinfo{author}{\bibfnamefont{S.}~\bibnamefont{Adam}},
  \bibinfo{author}{\bibfnamefont{E.~H.} \bibnamefont{Hwang}},
  \bibinfo{author}{\bibfnamefont{S.}~\bibnamefont{\mbox{Das Sarma}}},
  \bibinfo{author}{\bibfnamefont{H.~L.} \bibnamefont{Stormer}},
  \bibnamefont{and} \bibinfo{author}{\bibfnamefont{P.}~\bibnamefont{Kim}},
  \bibinfo{journal}{Phys. Rev. Lett.} \textbf{\bibinfo{volume}{99}},
  \bibinfo{pages}{246803} (\bibinfo{year}{2007}).

\bibitem[{\citenamefont{Rossi and \mbox{Das Sarma}}(2008)}]{kn:rossi2008}
\bibinfo{author}{\bibfnamefont{E.}~\bibnamefont{Rossi}} \bibnamefont{and}
  \bibinfo{author}{\bibfnamefont{S.}~\bibnamefont{\mbox{Das Sarma}}},
  \bibinfo{journal}{Preprint}  (\bibinfo{year}{2008}).

\bibitem[{\citenamefont{Martin et~al.}(2008)\citenamefont{Martin, Akerman,
  Ulbricht, Lohmann, Smet, \mbox{von Klitzing}, and Yacobi}}]{kn:martin2007}
\bibinfo{author}{\bibfnamefont{J.}~\bibnamefont{Martin}},
  \bibinfo{author}{\bibfnamefont{N.}~\bibnamefont{Akerman}},
  \bibinfo{author}{\bibfnamefont{G.}~\bibnamefont{Ulbricht}},
  \bibinfo{author}{\bibfnamefont{T.}~\bibnamefont{Lohmann}},
  \bibinfo{author}{\bibfnamefont{J.~H.} \bibnamefont{Smet}},
  \bibinfo{author}{\bibfnamefont{K.}~\bibnamefont{\mbox{von Klitzing}}},
  \bibnamefont{and} \bibinfo{author}{\bibfnamefont{A.}~\bibnamefont{Yacobi}},
  \bibinfo{journal}{Nature Physics} \textbf{\bibinfo{volume}{4}},
  \bibinfo{pages}{144} (\bibinfo{year}{2008}).

\bibitem[{\citenamefont{Chen et~al.}(2008)\citenamefont{Chen, Jang, Adam,
  Fuhrer, Williams, and Ishigami}}]{kn:chen2007b}
\bibinfo{author}{\bibfnamefont{J.~H.} \bibnamefont{Chen}},
  \bibinfo{author}{\bibfnamefont{C.}~\bibnamefont{Jang}},
  \bibinfo{author}{\bibfnamefont{S.}~\bibnamefont{Adam}},
  \bibinfo{author}{\bibfnamefont{M.~S.} \bibnamefont{Fuhrer}},
  \bibinfo{author}{\bibfnamefont{E.~D.} \bibnamefont{Williams}},
  \bibnamefont{and} \bibinfo{author}{\bibfnamefont{M.}~\bibnamefont{Ishigami}},
  \bibinfo{journal}{Nature Physics, in press (arXiv:0708.2408v1)}
  (\bibinfo{year}{2008}).

\bibitem[{\citenamefont{Han et~al.}(2007)\citenamefont{Han, Ozyilmaz, Zhang,
  and Kim}}]{kn:han2007}
\bibinfo{author}{\bibfnamefont{M.~Y.} \bibnamefont{Han}},
  \bibinfo{author}{\bibfnamefont{B.}~\bibnamefont{Ozyilmaz}},
  \bibinfo{author}{\bibfnamefont{Y.}~\bibnamefont{Zhang}}, \bibnamefont{and}
  \bibinfo{author}{\bibfnamefont{P.}~\bibnamefont{Kim}},
  \bibinfo{journal}{Phys. Rev. Lett.} \textbf{\bibinfo{volume}{98}},
  \bibinfo{eid}{206805} (\bibinfo{year}{2007}).

\bibitem[{\citenamefont{Miao et~al.}(2007)\citenamefont{Miao, Wijeratne, Zhang,
  Coskun, Bao, and Lau}}]{kn:miao2007}
\bibinfo{author}{\bibfnamefont{F.}~\bibnamefont{Miao}},
  \bibinfo{author}{\bibfnamefont{S.}~\bibnamefont{Wijeratne}},
  \bibinfo{author}{\bibfnamefont{Y.}~\bibnamefont{Zhang}},
  \bibinfo{author}{\bibfnamefont{U.}~\bibnamefont{Coskun}},
  \bibinfo{author}{\bibfnamefont{W.}~\bibnamefont{Bao}}, \bibnamefont{and}
  \bibinfo{author}{\bibfnamefont{C.}~\bibnamefont{Lau}},
  \bibinfo{journal}{Science} \textbf{\bibinfo{volume}{317}},
  \bibinfo{pages}{1530} (\bibinfo{year}{2007}).

\end{thebibliography}

\end{document}